\documentclass[journal]{IEEEtran}
\usepackage{nicematrix}
\usepackage{color}
\usepackage{cite}
\usepackage{textcomp}
\usepackage{xcolor}
\usepackage{balance}
\usepackage{graphicx}
\usepackage[english]{babel}
\usepackage{amsthm,amssymb}
\usepackage{amsmath}
\usepackage{subfigure}
\usepackage{mathtools}
\usepackage{tikz}
\usepackage{pgf}
\usepackage{amsfonts}
\usepackage{bm}
\usepackage[bottom]{footmisc}
\usepackage[ruled,linesnumbered]{algorithm2e}
\usepackage[T1]{fontenc}
\usepackage{bbm}
\usepackage{enumerate}

\usepackage{setspace}
\usepackage{array}
\usepackage{multirow}
\usepackage{diagbox}
\newtheorem{definition}{Definition}[section]

\newtheorem{theo}{Theorem}
\DeclareMathOperator*{\argmin}{arg\,min}
\newtheorem{rem}{Remark}

\allowdisplaybreaks
 
\def\BibTeX{{\rm B\kern-.05em{\sc i\kern-.025em b}\kern-.08em
    T\kern-.1667em\lower.7ex\hbox{E}\kern-.125emX}}
\begin{document}

\title{DRAMS: Double-RIS Assisted Multihop Routing Scheme for Device-to-Device Communication}

\author{Authors}
\author{Lakshmikanta Sau, Priyadarshi Mukherjee, and Sasthi~C.~Ghosh
\thanks{L. Sau, P. Mukherjee, and S. C. Ghosh are with the Advanced Computing \& Microelectronics Unit,  Indian Statistical Institute, Kolkata 700108, India. (E-mail: lakshmikanta030@gmail.com, priyadarshi@ieee.org, sasthi@isical.ac.in).
}}

\maketitle
\begin{abstract}
Reconfigurable intelligent surfaces (RISs) is a promising solution for enhancing the performance of multihop wireless communication networks. In this paper, we propose a double-RIS assisted multihop routing scheme for a device-to-device (D2D) communication network. Specifically, the scheme is dependent on the already deployed RISs and users in the surroundings. Besides the RISs, the emphasis of this work is to make more use of the existing intermediate users (IUs), which can act as relays. Hence, the density of RIS deployment in the surroundings can be reduced, which leads to the avoidance of resource wastage. However, we cannot solely depend on the IUs because this implies complete dependence on their availability for relaying and as a result, the aspect of reliability in terms of delay-constrained information transfer cannot be guaranteed. Moreover, the IUs are considered capable of energy harvesting and as a result, they do not waste their own energy in the process of volunteering to act as a relay for other users. Numerical results demonstrate the advantage of the proposed scheme over some existing approaches and lastly, useful insights related to the scheme design are also drawn, where we characterize the maximum acceptable delay at each hop under different set-ups.
\end{abstract}


\section{Introduction}
In the recent years, the wireless traffic has been increasing at an explosive rate; it is expected to increase more than five times in between $2023$ and $2028$ \cite{ericsson}. To support this need of enhanced data traffic, technologies such as beamforming and adaptive modulation have been developed over the last few decades. However, irrespective of the technicalities, the unifying motivation behind all of them is to intelligently adapt to the randomly varying wireless channel instead of having a control over it. In this context, a new technology that promises to address this issue is the so-called reconfigurable intelligent surfaces (RISs) \cite{risi1}. A RIS, consisting of an array of reconfigurable passive elements embedded on a flat metasurface, is able to `control' the wireless channel instead of adapting to it \cite{risi2}. This is practically done by tuning the parameters of its passive elements \cite{impl}. Furthermore, as a RIS simply reflects the incident signal in a desired direction, it does not need any radio-frequency (RF) chains. As a result, this reduces the hardware cost thereby enhancing the energy efficiency of the future wireless networks.

Motivated by this, the aspect of RIS assisted device-to-device (D2D) communications \cite{risi4,risi3} form an interesting direction of research. The work in \cite{risi4} investigates the role of RISs for enhancing the energy efficiency of a D2D communication network. The authors in \cite{risi3} focus on the uplink of a RIS assisted D2D enabled cellular networks. Moreover, the objective of obtaining high speed data rates is efficiently fulfilled by the use of high frequency signals, such as the millimeter waves (mmWaves) \cite{mm10}, for short distance communication. However, mmWaves suffer from its own set of shortcomings like significantly high penetration and propagation losses. Thus, RIS assisted D2D network is the solution for such scenarios, where the direct line of sight (LoS) is not of sufficient quality to support mmWave-based communication \cite{mmris1}. As a result, RISs are strategically placed at locations where they have clear LoS links with both the users intending to communicate with each other. In this context, the authors in \cite{opris,deb2021ris,sauris,dar} investigate the aspect of strategic RIS placement. Furthermore, \cite{opris,deb2021ris,sauris} consider a primary reflection-based single RIS system, i.e., the signal from a given user reaches its desired counterpart on being reflected by a single RIS placed within its communication range. As a result, the number of strategic RIS locations obtained in a particular environment is significantly large. On the other hand, the work in \cite{dar} involves only double reflection based scenario.

Accordingly, the work in \cite{dar,2hop} demonstrate that by proper tuning of the RISs, the multi-RIS secondary reflection can be leveraged to significantly enhance the range of communication. It is also to be noted that RISs are essentially passive devices, i.e., they simply reflect the incoming signal to a desired direction by tuning its parameters. As any given pair of users always communicate over a practically finite amount of time, having RISs deployed for all such potential pairs leads to unnecessary wastage of resources. A potential solution to avoid this wastage is a cooperative multihop framework \cite{mhop1,mhop2}. In other words, apart from using the RISs, the other users present in the surroundings, if they are idle, may act as a relay, namely amplify and forward (AF) or decode and forward (DF) \cite{relay10}, to facilitate the communication between a given pair of users. While AF relays facilitate low cost processing, they also result in boosting the effective noise at the desired user. On the other hand, DF relays guarantee high reliability as they forward only the received information and not the entire information-plus-noise mixture to the next hop \cite{dfaf10}. Thus, it appears that for the intermediate idle users, opting to act as DF relay is a beneficial solution. Note that, we cannot rule out the usage of RISs completely. If the communication between a pair of users is solely dependant on the intermediate users, the aspect of reliability, in terms of delay-constrained information transfer, cannot be guaranteed. In that case, the time for the entire communication process will be entirely dependant on the traffic characteristics of these intermediate users.

Moreover, it may apparently appear that the users, which agree to act as relays, are doing so by depleting their own energy source. In this context, we consider the green coexistence paradigm, i.e., all the users are equipped with an energy harvesting (EH) unit and if they agree to act as a relay, they do harvest energy from the received signal. This harvested energy can be interpreted as some `reward' for volunteering to act as relay in their idle time. By considering an over-simplified linear EH model instead of the actual non-linear one \cite{hmod}, the work in \cite{eh10} propose a similar approach in the context of wireless sensor networks. Furthermore, these works do not consider the impact of the user traffic characteristics, which in turn, is responsible for the EH time interval. Motivated by this, we consider the aspect of multi-RIS secondary reflection to look into a RIS-assisted multihop D2D framework for wireless networks, where the users are capable of harvesting energy, while acting as DF relays depending on their availability. Given a pair of devices that wish to communicate, we aim to find a sequence of RISs and IUs acting as DF relays in their idle time to establish communication when a direct link cannot be established between them. To the best of our knowledge, this is the first work that proposes such a multihop routing scheme, which we call double-RIS assisted multihop routing scheme (DRAMS), for a D2D communication network.

Specifically, DRAMS is based on the already deployed RISs and the users in the surroundings. It describes the procedure by which information transfer takes place from a particular user to its desired counterpart. Moreover, we assume that the RISs are strategically placed in the environment \cite{opris,deb2021ris} and the idle intermediate users (IUs) agree to act as a DF relay node. Our priority is to make more use of the idle IUs over the RISs due to the following reasons. Firstly, the RISs reflect the entire incoming signal, i.e., including the noise, in the desired direction whereas a DF relay separates the noise from the information to transmit only the latter to the next user. Secondly, being a passive device, too many RISs installation in the surroundings leads to unnecessary wastage of resources. Lastly, opting for RIS over an idle IU as a hop implies frequent restart of the former, which creates problems for other users that are being served by this particular RIS at that time. However, the importance of the RISs cannot be ruled out completely. Being solely dependent on the IUs for information transfer may also hamper the reliability of the entire process, as it will fully rely on the IU availability and traffic characteristics. Useful insights related to the proposed routing scheme are also obtained in this work, where we characterize the maximum acceptable delay under different scenarios. Finally, the numerical results demonstrate the benefit of DRAMS in terms of reduced RIS usage, enhanced data rate, and energy efficiency, respectively, with respect to the existing benchmark scheme.

The rest of this paper is as follows: Section II describes the system model and the problem formulation, Section III presents the proposed strategy, and Section IV analyses the delay associated with information transfer. Numerical results are presented in Section V and Section VI concludes the work.

\section{System Model}

\subsection{Network topology}
\begin{figure}
    \centering
    \includegraphics[width=\linewidth]{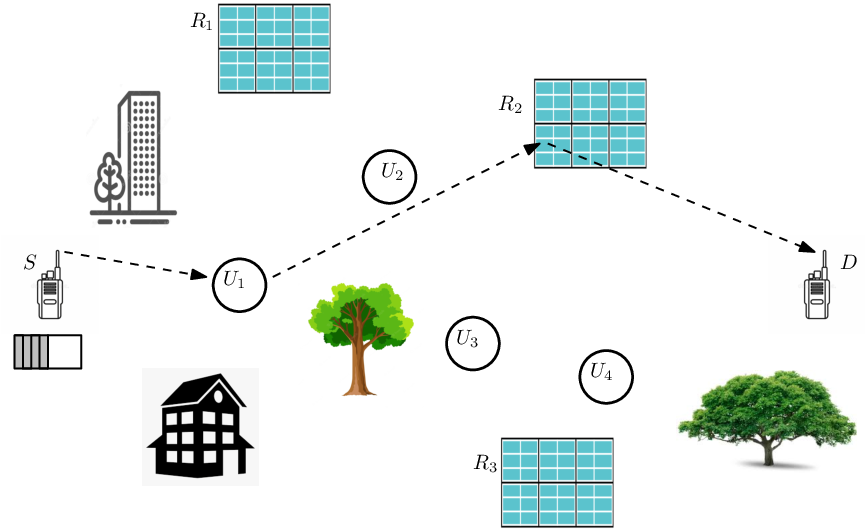}
        \vspace{-1mm}
    \caption{The proposed RIS-based multihop network architecture.}
     \label{system model}
    \vspace{-1mm}
\end{figure}

A wireless network topology is considered, which consists of a source $S$, $K$ RISs $R_1,R_2,\cdots,R_K$, $M$ IUs $U_1,U_2,$ $\cdots,U_M$, and destination $D$\footnote{Network topology with multiple $S-D$ pairs can be also considered, which is left for future work.}, respectively. Each transmitter, i.e., the source $S$ or an IU $U_j$ $j=1,\cdots,M$  transmits with the same fixed power $P$ and the RIS $R_i$ $i=1,\cdots,K$ has $N_i$ reflecting elements, respectively. Table \ref{notation} shows the list of variables used, along with the descriptions. In general, each RIS is effectively controlled to adjust both the amplitude and phase of the incident waveform. However, for the sake of simplicity and mathematical tractability, the amplitude factor is set to unity and it is only the phase that is tuned or optimized \cite{ris1,ris2}. Moreover, we assume that no direct LoS exists between $S$ and $D$ and that $S$ relies on the IUs and/or RISs to communicate to $D$. Each IU $U_j$ communicates with its own associated receiver and acts as a DF relay for other users, when idle. Lastly, each IU is equipped with a buffer of sufficient capacity and all the D2D pairs follow time-slotted synchronous communication \cite{sync}, with slot duration $T_s$. An example of this topology is presented in Fig. \ref{system model}, for $K=3$ and $M=4$, respectively, where $S$ can communicate with $D$ via IUs and/or RISs.

{\color{red}{\begin{table*}[!t]
\centering
 \caption{SUMMARY OF NOTATIONS.} \label{notation}
\resizebox{0.8\textwidth}{!}{%
  \begin{tabular}{|c|c||c|c|}
    \hline \hline
    \textbf{Notation} & \textbf{Description} &  \textbf{Notation} &  \textbf{Description}\\
    \hline
     $K,M$ & Number of RISs and IUs respectively & $T_s$ & Slot duration\\
    \hline
     $N_i$ & Number of elements in $i$-th RIS & $R_i$ & $i$-th RIS \\
    \hline
   $\theta_n$ & phase shift at $n$-th RIS element & $\lambda_k(\mu_k)$ & ON(OFF) period length \\
    \hline
    $\rho_L$ & Path-loss at one meter distance & $\alpha$ & Path-loss exponent\\
    \hline
    $d_{m,n}$ & Distance between nodes $m$ and $n$ & $h$ & Complex channel gain \\

    \hline
	$\Phi_i$ & Phase shift matrix for $i$-th RIS & $U_i$ & $i$-th IU\\
    \hline
    $P$, $P_{\rm harv},$ and $P_{\rm proc}$ & Transmit , harvested, and processing power respectively & $M_h$ & Maximum harvested power\\
    \hline
	$l$ & Euclidean distance from $S$ to $D$ & $\Omega_I(\Omega_B)$ & Set of idle (busy) IUs\\   
    \hline
    $t_i$ & Actual waiting time at $U_i$ & $m_q$ & Modulation scheme \\
    \hline
    $\nu_I(\nu_B)$ & Duration of idleness (busyness) & $\alpha_k$ & Number of packets \\
    \hline
    $\tau_{\rm req}$ & Required slots for information transfer & $E_{\rm req}$ & Required energy\\
    \hline
    $\eta_w$ & Waiting time (in slots) at a busy IU  & $P_b$ & Bit error rate\\
    \hline
    $\psi$ & Minimum number of hops for information transfer & $T_d$ & Total delay\\
    \hline
    $T_{d_i}$ &  Maximum waiting time at $i$-th IU & $\gamma$ & SINR\\
    \hline
    $P_{\rm th}$ & Acceptable threshold probability & $R$ & Achievable data rate \\
    \hline
    $\chi$ & Number of hops between $S$ and $D$ & $\mathcal{D_T}$ & Data throughput\\
    \hline
     $\sigma^2$ & Variance of the AWGN & $\mathcal{E}_{\rm eff}$ & Energy efficiency\\
    \hline
    $\delta$ & Acceptable error threshold & $r$ & Coverage area\\
    \hline
  \end{tabular}
  }
  \vspace{-3.2mm}
\end{table*}}}

\subsection{User Traffic Characterization} \label{tcharc}
It is noted that in a typical wireless communication scenario, data generally arrives in bursts to the users. As a result, the IUs $U_1,U_2,\cdots,U_M$ in this work are characterized by exponentially distributed OFF and ON period lengths, with means $\lambda_k$ and $\mu_k$, respectively. Without any loss of generality, $T_s$ is assumed to be small in comparison with $\mu_k$ and $\lambda_k$ \cite{crn}, which prevents $U_k$ $\forall$ $k=1,\cdots,M$ changing its status multiple times within a single $T_s$. Thus, the state $0$ and $1$ represents the IU being idle and busy, respectively and the probability $p_{ij}$ $\forall$ $i,j \in \{0,1\}$ denotes the transition probability of the IU currently being in state $i$ and it changes to state $j$ in the next time slot. Accordingly, the IU activities are characterized by a discrete-time Markov chain (DTMC) with the state transition probabilities \cite{crn}:
\vspace{-1mm}
\begin{equation}
\text{p}_{10}=\int_0^{T_s}\frac{1}{\mu_k}e^{-a/\mu_k}da=1-e^{-T_s/\mu_k}, \:\: \text{p}_{11}=1-\text{p}_{10},
\end{equation}
\begin{equation*}
\text{p}_{01}=\int_0^{T_s}\frac{1}{\lambda_k}e^{-b/\lambda_k}db=1-e^{-T_s/\lambda_k}, \:\: \text{and} \:\: \text{p}_{00}=1-\text{p}_{01}. 
\end{equation*}
Accordingly the state transition matrix $\mathcal{P}$ is:
\vspace{-1mm}
\begin{equation} \label{tpmtrx}
\mathcal{P}=
\begin{bmatrix}
   \text{p}_{00} & \text{p}_{01} \\
   \text{p}_{10} & \text{p}_{11} 
\end{bmatrix}=
 \begin{bmatrix}
   e^{-T_s/\lambda_k} & 1-e^{-T_s/\lambda_k} \\
   1-e^{-T_s/\mu_k} & e^{-T_s/\mu_k}
\end{bmatrix}.
\end{equation}

\subsection{Channel Model}  \label{cmod}
Depending on the availability of the IUs, it is possible to connect $S$ to $D$ with or without taking the help of any RIS. When RIS is being used, the signal from an arbitrary $U_i$ can reach $U_{i+1}$ via any of the following paths: (i) primary reflection, i.e., the signal from $U_i$ reaches $U_{i+1}$ by using only one RIS and (ii) secondary reflection, i.e., there exist two consecutive RISs via which the signal from $U_i$ reaches $U_{i+1}$. In this work, due to large effective path loss, we neglect the aspect of triple or higher order reflections \cite{2hop}. However, the secondary reflection are not negligible in practice, especially in urban environments, where the RISs are not deployed too far from each other. This problem can be modeled as a graph, where the devices represent the vertices. These vertices have an edge in between if and only if the corresponding nodes can communicate, i.e., they reside within some threshold distance. We assume that the wireless links suffer from both large-scale path-loss effects and small-scale block fading. The channels $S \rightarrow U_j$, $U_j \rightarrow D$, and $U_j \rightarrow U_k$ $\forall$ $j,k=1,\cdots,   M$ exhibit small-scale fading and their corresponding path-loss factors are $\rho_{\rm L}^{1/2}d_{SU_j}^{-\alpha/2},\rho_{\rm L}^{1/2}d_{U_jD}^{-\alpha/2},$ and $\rho_{\rm L}^{1/2}d_{U_jU_k}^{-\alpha/2}$, respectively, where $\rho_{\rm L}$ is the pathloss at one meter distance, $\alpha$ is the path-loss exponent and $d_{mn}$ denotes the distance between $m$ and $n$.

Let $\boldsymbol{h}_{S/UR_i} \in \mathbb{C}^{N\times 1}$, $\boldsymbol{h}_{R_iR_j} \in \mathbb{C}^{N\times N}$, and $\boldsymbol{h}_{R_jD/U} \in \mathbb{C}^{1\times N}$ denote the channel matrix from $S$ or IU to $i$-th RIS, $i$-th to $j$-th RIS $(i \neq j)$, and $j$-th RIS to an IU or $D$, respectively. In addition, the phase-shift matrix of the $i$-th RIS is denoted by $\boldsymbol{\Phi}_i={\rm diag}(\phi_1,\cdots,\phi_N) \in \mathbb{C}^{N \times N}$, i.e., a diagonal matrix accounting for the response of the RIS elements, where $\phi_n=\exp(j\theta_n),$ $n=1,\cdots,N$ and $\theta_n \in [0,2\pi]$ is the phase shift applied by the RIS elements \cite{phin}. Lastly, the total path-loss for each of these channel matrix is the product of the path-loss of each point-to-point link \cite{2hop}. 
Accordingly, the effective channel gain in case of single and double reflection is $\boldsymbol{h}_{R_iD/U}\boldsymbol{\Phi}_i\boldsymbol{h}_{S/UR_i}$ and $\boldsymbol{h}_{R_jD/U}\boldsymbol{\Phi}_j\boldsymbol{h}_{R_iR_j}\boldsymbol{\Phi}_i\boldsymbol{h}_{S/UR_i}$, respectively.

\subsection{Energy Harvesting Model}
As stated earlier, it is for an idle IU $U_j$ $\forall$ $j=1,\cdots,M$ to decide whether to act as a DF relay or not. Moreover, there must be some `reward' for the same or else, there is no point for the user to waste its own energy in transferring data packets from $S$ to $D$. In this context, we assume that each $U_j$ is equipped with an EH unit, which can extract DC power from the received electromagnetic waves \cite{harvm}. If an idle $U_j$ agrees to act as a relay, we incentivize it in the form of a reward, i.e., it is able to harvest energy from the incoming signal and use the same to transfer the received information. For a transmission power $P$, the power harvested at IU is \cite{hmod}
\vspace{-1mm}
\begin{equation} \label{harv}
P_{\rm harv}=\frac{M_h(1-e^{-aP\rho_{\rm L}d^{-\alpha}|h|^2})}{1+e^{-a(P\rho_{\rm L}d^{-\alpha}|h|^2-b)}},
\end{equation}
where $M_h$ is the maximum harvested power corresponding to the saturated EH circuit, $h$ is the complex channel gain, $d$ is the associated distance, and finally, $a$ and $b$ are the respective circuit parameters.

\subsection{Delay-constrained Transmission}
Shannon capacity is the largest data rate at which the information can be transmitted with an arbitrarily small error probability, provided that the number of channel uses is infinitely large \cite{shanc}. However, for applications such as delay-constrained scenarios, the number of channel uses cannot be very large. As a result, the error probability will not be arbitrarily small and it needs reconsideration. In such scenarios,  the maximum instantaneous achievable data rate $R$ is approximated as \cite{shanf}
\begin{equation} \label{rate}
    R(\gamma)=\log_2 (1+\gamma)-\frac{Q^{-1}(\varepsilon)}{\ln 2} \sqrt{\frac{\gamma^2+2\gamma}{M_b(1+\gamma)^2}},
\end{equation}
where $\gamma$ is the signal to noise ratio (SNR), $\varepsilon \in [0,1]$ is the error probability, $M_b$ is the number of channel uses, and $\displaystyle Q(x)=\frac{1}{\sqrt{2\pi}}\int\limits_{x}^{\infty}e^{-\frac{t^2}{2}}dt$ is the Gaussian $Q$ function. For delay unconstrained scenarios, i.e., when $M_b \rightarrow \infty$, we have $R(\gamma) \rightarrow \log_2 (1+\gamma)$. When a RIS is selected to pass the signal due to unavailability of idle IUs, we consider this achievable data rate $R$ while searching for an IU in the next hop.

\section{DRAMS: The Proposed Strategy}

This section discusses the proposed multihop framework DRAMS in detail, where the novelty lies in the joint IU traffic characteristics and double-RIS assisted dynamic framework. As we are considering a delay-constrained scenario, the data from $S$ must reach $D$ within time $T_d$ in this set-up. Here we assume that a device cannot communicate with another beyond a distance $r$ and $S$ has $\alpha$ packets of information to send $D$ with $\varphi$ bits in each. Pictorially, we connect the location of $S$ and $D$ by a virtual straight line and consider it to be the x-axis. Accordingly, we consider another virtual line as the y-axis at $S$, which is perpendicular to the x-axis. We intend to connect $S$ to $D$ via some IUs/RISs. Firstly, we scan the right half circle\footnote{It is to be noted that in our proposed framework, we search for appropriate IU/RIS in the right half circle. The reason for this is attributed to the fact that we have assumed to have $D$ in the right-hand side of $S$. However, if $D$ happens to be on the left-hand side, the framework still works and it is just that we then search for appropriate IU/RIS in the left-hand circle at each hop. By generalizing this, it can be said that we search for IU/RIS in the direction of $D$ from $S$.} of radius $r$  at $S$ to identify the idle IUs. Secondly, after the idle IU identification, we decide on the appropriate modulation scheme and its corresponding energy requirement. Thirdly, in case of multiple idle IU availability, we chose the appropriate IU based on the least remaining distance (LRD) from $D$ and the acceptable delay constraint.  Two IUs are said to be directly connected if there exists a LoS in between them and they reside within a distance $r$ of each other. Else, we identify suitable RISs for this purpose. Finally, we also provide an illustrative example of the proposed DRAMS. 

\subsection{Identification of the Idle IUs}
We identify the idle IUs by beacon transmission within a radius $r$, in the direction of $D$ \cite{beacon1}, which can act as potential DF relays. We define $\Omega=\{U_1,\cdots,U_{\epsilon} \}$ as the set of all $U_j$s that are present in the right half circle of radius $r$ centred at $S$, where $\epsilon<K$ and $U_k=1/0,$ depending on whether the $k^{\text{th}}$ IU is busy/idle. As we intend to reduce the LRD in each hop, we consider only the right half circle for identifying the potential relays. Accordingly, we define the set of idle/busy IUs as
\vspace{-1mm}
\begin{equation}  \label{bidef}
    \Omega_I=\{u_I^1,u_I^2,\cdots,u_I^{\epsilon_I}\}\:\:\text{and}\:\: \Omega_B=\{u_B^1,u_B^2,\cdots,u_B^{\epsilon_B}\},
\end{equation}
where $\epsilon_I+\epsilon_B=\epsilon$ denotes the total number of IUs in the concerned region. On the basis of the traffic characteristics of a particular idle (busy) IU, we estimate the time for which it continues to remain idle (busy) given that it is currently idle (busy).


\begin{definition} Duration of Idleness (DoI)
$\nu_I$: It is the time duration during which a particular IU is estimated to be idle, given that it is currently idle.
\end{definition}

\noindent From the transition probability matrix (\ref{tpmtrx}), we know that the 
$\text{p}_{00}$ corresponding to the idle IU $U_k$ is $\text{p}_{00}=e^{-T_s/\lambda_k}$. As we are considering exponentially distributed idle and busy periods, due to the memoryless property \cite{papoulis}, we obtain $\nu_I^k$, i.e., $\nu_I$ for the IU $U_k$, as follows. For an acceptable error threshold $\delta$, we desire to have
\vspace{-1mm}
\begin{equation} \label{zi}
\text{p}_{00}^{\nu_I^k} \geq 1-\delta \implies \nu_I^k \leq \frac{\lambda_k}{T_s}\ln \left( \frac{1}{1-\delta} \right).
\end{equation}
Since we are interested in considering the maximum time for which an IU is expected to be idle for a given $\delta$, we consider $\nu_I^k = \frac{\lambda_k}{T_s}\ln \left( \frac{1}{1-\delta} \right)$. Hence, corresponding to $\Omega_I$, we have $\varphi_I$, where $\varphi_I=\{\nu_I^1,\nu_I^2,\cdots,\nu_I^{\epsilon_I} \}$. Similarly, we define a metric `Duration of Busyness' (DoB).

\begin{definition} Duration of Busyness (DoB)
$\nu_B$: It is the time duration during which a particular IU is estimated to be busy, given that it is currently busy.
\end{definition}

\noindent By considering $\text{p}_{11}=e^{-T_s/\mu_k}$ from \eqref{tpmtrx} and adopting a similar procedure as DoI, the DoB $\nu_B^k$, i.e., $\nu_B$ for $U_k$, is obtained as
\vspace{-1mm}
\begin{equation} \label{zb}
\nu_B^k = \frac{\mu_k}{T_s}\ln \left( \frac{1}{1-\delta} \right).
\end{equation}
Hence, corresponding to $\Omega_B$, we obtain $\varphi_B=\{\nu_B^1,\nu_B^2,\cdots,\nu_B^{\epsilon_B} \}$.

\subsection{ Choosing Appropriate Modulation at the IUs}\label{modulation}
The objective of this work is not only to transfer data from $S$ to $D$, but also with minimum energy consumption. In the process of doing so, we introduce the aspect of rate adaptation at the IUs. Moreover, the idle IUs employ rate adaptation if and only if there is a direct connection between the users and there is no RIS used to connect them\footnote{The aspect of RIS-enabled rate adaptation is not considered here, as this is beyond the scope of the current work. However, the proposed framework can also be extended to such scenarios with suitable adjustments like optimal RIS beam alignment \cite{2hop}.}. The IUs adopt a
modulation scheme $m_q$ from the set $\mathbb{M}=\{m_1,m_2,\cdots,m_{|\mathbb{M}|}\}$, which corresponds to a data rate $D_q=\log_2(m_q)$ $\forall$ $q=1,\cdots,|\mathbb{M}|$. The choice of $m_q$ depends on the wireless channel between the two consecutive IUs, where the complex channel gain $h$, bit error rate (BER) $P_b$, constellation size $m_q$, and the received power $P\rho_{\rm L}d^{-\alpha}|h|^2$ are related as \cite{goldsmithwc}:
\vspace{-1mm}
\begin{equation} \label{mqam}
P_b=c_1 \exp \left( \frac{-c_2 P\rho_{\rm L}d^{-\alpha}|h|^2}{\sigma^2 (m_q^{c_3}-c_4)} \right).
\end{equation}
Here $\sigma^2$ is the noise power, and $c_1,\cdots,c_4$ are modulation-specific constants, respectively. The above equation explains the relation between the chosen modulation scheme and the application-specific acceptable BER. By considering a transmission power $P$ and a modulation scheme $m_q$, the time required for complete transfer of $\alpha_k$ information packets of the IU $U_k$ with $\varphi$ bits in each is $\displaystyle\tau_{\text{req}}(k,q)=\Bigl\lceil\frac{\alpha_k \varphi}{D_q}\Bigr\rceil$ slots, where $\Bigl\lceil\cdot\Bigr\rceil$ denotes the ceiling function. Accordingly, if $P_{\text{proc}}$ is the processing power, the corresponding energy required for complete information transfer is characterized as
\vspace{-1mm}
\begin{equation}  \label{enrgy}
E_{\text{req}}(k,q)=(P+P_{\text{proc}})\tau_{\text{req}}(k,q)T_s,
\end{equation}
which is not a fixed quantity, but a function of the chosen constellation size.

\subsection{Selection of the Appropriate IUs}
Till now, we have identified the pool of busy and idle IUs within the radius $r$ towards $D$. Now, among the idle ones, we identify the ones that can be leveraged upon to act as DF relays in forwarding the information from $S$ to $D$. Towards this direction, we define the set $\mathcal{Y}_{k,q}$ as
\begin{equation}  \label{ykq}
\mathcal{Y}_{k,q}= \bigl\{ \tau_{\text{req}}(k,q) : U_k \in \Omega_I, \:\: m_q \in \mathbb{M}, \:\: \tau_{\text{req}}(k,q) \leq \nu_I^k \bigr\}.
\end{equation}
Specifically, $\mathcal{Y}_{k,q}$ is the set of those IU $U_k \in \Omega_I$ and their associated modulation scheme $m_q \in \mathbb{M}$ such that they satisfy what we define as the \textit{availability constraint} $\tau_{\text{req}}(k,q) \leq \nu_I^k$, i.e., the required time for complete information transfer from $U_k$ is less than its estimated idle time $\nu_I^k$. Accordingly, $S$ chooses the appropriate IU $U_{k^*}$ along with its modulation scheme $m_{q^*}$ as
\begin{equation}  \label{optkq}
\left(k^*,q^*\right)=\argmin_{k,q} \mathcal{Y}_{k,q}.
\end{equation}
Note that the computational complexity of this entire process is $|\Omega_I|\times|\mathbb{M}|$.
However, at times we have $\mathcal{Y}_{k,q}$ to be a null set as $\tau_{\text{req}}(k,q) >  \nu_I^k(U_k)$ $\forall$ $U_k \in \Omega_I$, i.e., DoI corresponding to all the elements of $\Omega_I$ is less than the time required by $U_k$ for continuous complete information transfer by using a constellation of size $m_q$. As it is necessary to complete the entire information transfer in a single phase, $S$ decides to avoid transmission and waits for an interval of $\eta_w$ slots with the hope that some $U_k \in \Omega_B$ may become idle in the near future. In the following subsection, we derive the analytical expression for the quantity $\eta_w$.

\subsubsection{Calculation of $\eta_w$} \label{wtime}
In such scenarios, $S$ identifies the pool of busy IUs in the right half circle of radius $r$, i.e., $\Omega_B$ from \eqref{bidef}. Accordingly, it estimates the time interval of $\eta_{I,n}$ slots after which $U_n \in \Omega_B$ will become idle, i.e., we have $U_n \in \Omega_I$ after $\eta_{I,n}$ slots. This can be effectively modeled as a geometric distribution \cite{papoulis}, where we map the event of $U_n$ being idle and busy as success and failure, respectively. Therefore, given that $U_n \in \Omega_B$, we are interested in finding out the number of trial required till the first success, which in this case, is $\eta_{I,n}$. From the transition probability matrix $\mathcal{P}$ in \eqref{tpmtrx}, we obtain the probability of success in the $\eta_{I,n}$-th slot as $\text{p}_{11}^{\eta_{I,n}-1}\text{p}_{10}$, which needs to be  greater than an acceptable threshold probability $p_{\rm th}$, i.e., 
\vspace{-1mm}
\begin{align}
    & \text{p}_{11}^{\eta_{I,n}-1}\text{p}_{10} \geq p_{\rm th}  \nonumber \\
    & \implies 
    \left( e^{-T_s/\mu_k} \right)^{\eta_{I,n}-1} \left( 1-e^{-T_s/\mu_k} \right) \geq p_{\rm th},
\end{align}
which after some trivial algebraic manipulations yield
\vspace{-1mm}
\begin{equation} \label{iwait}
    \eta_{I,n} \leq \frac{\mu_k}{T_s} \ln \left( \left(\frac{1-e^{-T_s/\mu_k}}{p_{\rm th}} \right)^+\right) +1,
\end{equation}
where we define $x^+=\max(x,1)$. Since we are interested in maximizing the chance of an IU availability, for a given $p_{\rm th}$, we consider $\eta_{I,n}=\frac{\mu_k}{T_s} \ln \left( \left(\frac{1-e^{-T_s/\mu_k}}{p_{\rm th}} \right)^+\right) +1$. Furthermore, as this is a delay-constrained scenario, we cannot afford to wait for a significant amount of time. Hence, $S$ chooses not to transmit for $\eta_w$ slots, where
\vspace{-1mm}
\begin{equation}
    \eta_w= \min\limits_{U_n \in \Omega_B} \eta_{I,n}.
\end{equation}
Since $S$ waits for $\eta_w$ slots, we have the reduced delay bound  $T_d^{'}$, where $T_d^{'}=T_d- \eta_w T_s$. Hence, we can say that the acceptable delay bound becomes tighter with every transmission deferral. In this context, a detailed delay analysis is provided later in Section \ref{DA}. Now after a time interval of $\eta_w$ slots, $S$ again scans the right half circle of radius $r$ to classify the idle IUs and accordingly, it solves \eqref{optkq} again to obtain $U_{k^*}$ along with the appropriate modulation $m_{q^*}$. If even now, the solution is a null set, $S$ proceeds with a RIS as the next hop.

\begin{rem}
We adopt a cooperative framework in this work, i.e., a particular IU agrees to act as a DF relay whenever it is idle. A scenario, where an idle IU may not agree to act as a relay for a particular $S-D$ pair, in spite of being idle, is not considered here but left for future work.
\end{rem}

\subsubsection{Harvested Energy}
As stated earlier, we incentivize the selected IU in terms of harvested energy, i.e., $U_{k^*}$ harvests energy from the signal it received from $S$. From \eqref{harv}, when a particular constellation $m_q$ is selected, we obtain the harvested energy at $U_{k^*}$ as
\vspace{-1mm}
\begin{equation}  \label{hsu}
    E_{\rm harv}(k^*,q^*)=\frac{M_h(1-e^{-aP\rho_{\rm L}d_{SU}^{-\alpha}|h_{SU}|^2})}{1+e^{-a(P\rho_{\rm L}d_{SU}^{-\alpha}|h_{SU}|^2-b)}}\tau_{\rm req}(k^*,q^*),
\end{equation}
where $h_{SU}\sim \mathcal{CN}(0,1)$ is the channel gain and $d_{SU}^{-\alpha}$ is the corresponding path-loss factor. Moreover, we observe from \eqref{hsu} that the harvested energy is a function of the chosen constellation $m_q$. Furthermore, \eqref{optkq} chooses the IU with the best channel condition and as $E_{\rm harv}$ is a monotonic function of $|h_{SU}|^2$, this results in better EH performance at $U_{k^*}$.

\begin{figure}
    \centering
    \includegraphics[width=\linewidth]{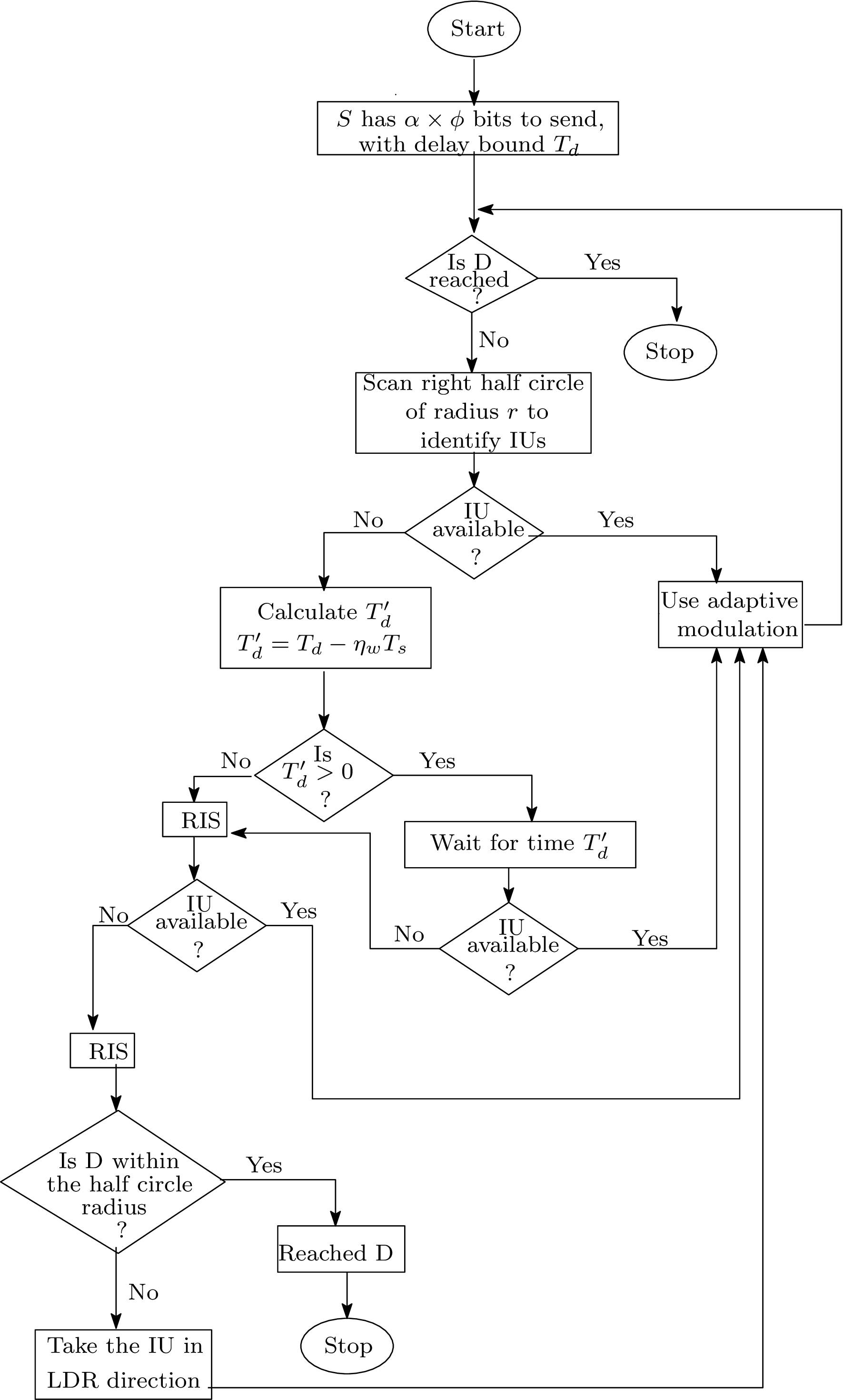}
    \vspace{-6mm}
    \caption{Flowchart of the Proposed Strategy}
    \label{flowchart}
    \vspace{-6mm}
\end{figure}

This entire procedure of identifying idle IUs within radius $r$ to act as relays is described later pictorially in Fig. \ref{flowchart}. However, if only IUs are to made act as relays, then the information transfer from $S$ to $D$ will solely rely on the IU traffic characteristics and activities. To overcome this problem, we take help of the already strategically placed RISs in the surroundings \cite{deb2021ris,sauris}, which always guarantees communication in the absence of appropriate IUs. This also implies that the location of the RISs are known to all the IUs in the surroundings. 

\subsection{Identification of RIS in case of Idle IU Unavailability}
In case of directly connected idle IU unavailability, $S$ searches for a RIS within the right half circle of radius $r$, which reduces the LRD for the next hop. The location of the idle IU in the proximity of the RIS is known to $S$ by the reverse path forwarding procedure \cite{beacon1} via the RIS, when an IU receives the beacon signal transmitted by $S$. In a multi-user scenario, by assuming that there are $L$ device pairs being supported by that particular RIS and there exists an idle IU in the right half circle of radius $r$ that satisfies the constraints as stated in \eqref{ykq}, the signal-to-interference-plus-noise-ratio (SINR) corresponding to this particular device pair is
\vspace{-1mm}
\begin{equation}
    \!\gamma_{\rm sr}\!=\!\frac{P|\boldsymbol{h}_{R_iD/U}\boldsymbol{\Phi}_i\boldsymbol{h}_{S/UR_i}|^2 d_{R_iD/U}^{-\alpha}d_{S/UR_i}^{-\alpha}\rho_{\rm L}^2}{\displaystyle\sum\limits_{\substack{l=1\\l \neq i}}^L P\rho_{\rm L}^2|\boldsymbol{h}_{R_lD/U}\boldsymbol{\Phi}_l\boldsymbol{h}_{S/UR_l}|^2d_{R_lD/U}^{-\alpha}d_{S/UR_l}^{-\alpha} + \sigma^2}, \label{sr}
\end{equation}
where $\sigma^2$ is the variance of the circularly symmetric zero mean additive white Gaussian noise (AWGN), the effective channel gain between $S$ and the idle IU is $\boldsymbol{h}_{R_iD/U}\boldsymbol{\Phi}_i\boldsymbol{h}_{S/UR_i}$ and, $\rho_{\rm L}^2d_{R_iD/U}^{-\alpha}d_{S/UR_i}^{-\alpha}$ is the effective path-loss factor as defined in Section \ref{cmod}. Accordingly, the phase shift matrix of this particular RIS is optimized, such that the sum throughput of all the users being served by the RIS is maximized.  Moreover, note that we are considering a delay-constrained scenario here, i.e., the data rate that we aim to maximize is obtained by replacing $\gamma=\gamma_{\rm sr}$ in \eqref{rate} as
\vspace{-1mm}
\begin{equation}
    R(\gamma_{\rm sr})=\log_2 (1+\gamma_{\rm sr})-\frac{Q^{-1}(\varepsilon)}{\ln 2} \sqrt{\frac{\gamma_{\rm sr}^2+2\gamma_{\rm sr}}{M_b(1+\gamma_{\rm sr})^2}},
\end{equation}
where $M_b$ denoting the total number of channel uses, is a finite quantity and $\varepsilon$ is the acceptable probability of error.

For the scenario of a single-user system, \eqref{sr} reduces to
\vspace{-1mm}
\begin{equation}
    \gamma_{\rm sr}=\frac{P|\boldsymbol{h}_{R_iD/U}\boldsymbol{\Phi}_i\boldsymbol{h}_{S/UR_i}|^2 \rho_{\rm L}^2d_{R_iD/U}^{-\alpha}d_{S/UR_i}^{-\alpha}}{\sigma^2},
\end{equation}
where we have $\boldsymbol{h}_{R_iD/U}=\Bigl[ \zeta_1e^{-j\theta_{\zeta,1}}, \cdots,\zeta_Ne^{-j\theta_{\zeta,N}} \Bigr],$ $\boldsymbol{h}_{S/UR_i}=\Bigl[ \omega_1e^{-j\theta_{\omega,1}}, \cdots,\omega_Ne^{-j\theta_{\omega,N}} \Bigr]$, and $\boldsymbol{\Phi}_i={\rm diag}(\phi_1,\cdots,\phi_N)$ with $\phi_n=\exp(j\theta_n),$ $n=1,\cdots,N$. Here, the optimal choice of $\boldsymbol{\Phi}_i$ which maximizes $\gamma_{\rm sr}$ is
\vspace{-1mm}
\begin{equation}  \label{optsu}
    \phi_n=\theta_{\zeta,n}+\theta_{\omega,n} \:\: \forall \:\: n=1,\cdots,N
\end{equation} and accordingly, we obtain
\vspace{-1mm}
\begin{equation}  \label{1user}
    \gamma_{\rm sr}^{\rm opt}=\dfrac{P \left( \sum\limits_{i=1}^N \zeta_i\omega_i \right)^2 \rho_{\rm L}^2 d_{R_iD/U}^{-\alpha}d_{S/UR_i}^{-\alpha}}{\sigma^2}.
\end{equation}
Furthermore, the received power at the IU  can be expressed as
$P|\boldsymbol{h}_{R_iD/U}\boldsymbol{\Phi}_i\boldsymbol{h}_{S/UR_i}|^2 \rho_{\rm L}^2d_{R_iD/U}^{-\alpha}d_{S/UR_i}^{-\alpha}$ and accordingly, the harvested power, in this case, is obtained from \eqref{harv}, i.e.,
\vspace{-1mm}
\begin{equation}
    P_{\rm harv}=\frac{M_h(1-e^{-aP|\boldsymbol{h}_{R_iD/U}\boldsymbol{\Phi}_i\boldsymbol{h}_{S/UR_i}|^2 \rho_{\rm L}^2d_{R_iD/U}^{-\alpha}d_{S/UR_i}^{-\alpha}})}{1+e^{-a(P|\boldsymbol{h}_{R_iD/U}\boldsymbol{\Phi}_i\boldsymbol{h}_{S/UR_i}|^2 \rho_{\rm L}^2d_{R_iD/U}^{-\alpha}d_{S/UR_i}^{-\alpha}-b)}},
\end{equation}
which can be further analytically characterized if the probability distribution function of $\boldsymbol{h}_{R_iD/U}$ and $\boldsymbol{h}_{S/UR_i}$ are known \cite{harvd}. Finally, based on \eqref{1user}, the maximum harvested power at the IU for a single user system is
\vspace{-1mm}
\begin{equation}
    P_{\rm harv}^{\rm opt}=\frac{M_h(1-e^{-aP \left( \sum\limits_{i=1}^N \zeta_i\omega_i \right)^2\rho_{\rm L}^2 d_{R_iD/U}^{-\alpha}d_{S/UR_i}^{-\alpha}})}{1+e^{-a(P \left( \sum\limits_{i=1}^N \zeta_i\omega_i \right)^2 \rho_{\rm L}^2d_{R_iD/U}^{-\alpha}d_{S/UR_i}^{-\alpha}-b)}}.
\end{equation}

\begin{rem}  \label{rem2}
The optimal phase shift matrix for a single-user system is obtained in \eqref{optsu}. However, finding the same for the scenario involving an arbitrary number of users is not as straightforward as above and the said problem has been formulated as an optimization problem in 
 \cite{risi3,2hop,ris2}.
\end{rem}

Note that it may happen as to there is no idle IU in the right half circle of radius $r$ centred at the RIS, which satisfies the desired criteria as mentioned in \eqref{ykq}. However, we are sure to find another RIS, as we assume that the RISs are already strategically placed in the surroundings \cite{deb2021ris}. Hence, the RIS directs the signal towards this newly found RIS in its range. Finally, if there is no idle IU in the coverage area, i.e., a right half circle of radius $r$, from this RIS, the communication stops. However, if an idle IU exists, the received power at this idle IU is
\vspace{-1mm}
\begin{align}      \label{pdr}
P_{\rm dr} & = P|\boldsymbol{h}_{R_jD/U}\boldsymbol{\Phi}_j\boldsymbol{h}_{R_iR_j}\boldsymbol{\Phi}_i\boldsymbol{h}_{S/UR_i}|^2 \nonumber \\
& \times\rho_{\rm L}^3d_{R_jD/U}^{-\alpha}d_{S/UR_i}^{-\alpha}d_{R_iR_j}^{-\alpha}
\end{align}
and by assuming that there are $L$ D2D pairs being served by this RIS at this point of time, the resulting doubly reflected SINR in this case is evaluated as
\vspace{-1mm}
\begin{equation}
    \gamma_{\rm dr}=\frac{\splitfrac{P|\boldsymbol{h}_{R_jD/U}\boldsymbol{\Phi}_j\boldsymbol{h}_{R_iR_j}\boldsymbol{\Phi}_i\boldsymbol{h}_{S/UR_i}|^2}{\times \rho_{\rm L}^3d_{R_jD/U}^{-\alpha}d_{S/UR_i}^{-\alpha}d_{R_iR_j}^{-\alpha}}}{\displaystyle\sum\limits_{\substack{l=1\\l \neq i}}^L \splitfrac{P|\boldsymbol{h}_{R_jD/U}\boldsymbol{\Phi}_j\boldsymbol{h}_{R_lR_j}\boldsymbol{\Phi}_l\boldsymbol{h}_{S/UR_l}|^2}{\times \rho_{\rm L}^3d_{R_jD/U}^{-\alpha}d_{S/UR_l}^{-\alpha}d_{R_lR_j}^{-\alpha} }+ \sigma^2}.\label{dr}
\end{equation}
Our aim is to maximize $R(\gamma_{\rm dr})$, which is obtained by replacing $\gamma=\gamma_{\rm dr}$ in \eqref{rate}. As stated earlier in Remark \ref{rem2}, we optimize the phase shift matrix by using any of the existing available techniques. Lastly, the harvested power, in this context of a doubly RIS reflected signal, is obtained from \eqref{harv} as 
\vspace{-1mm}
\begin{equation}
    P_{\rm harv}=\frac{M_h(1-e^{-aP_{\rm dr}})}{1+e^{-a(P_{\rm dr}-b)}},
\end{equation}
where $P_{\rm dr}$ is the received power as defined in \eqref{pdr}. Once the information reaches an idle IU in one of the hops, then the same process, as described above, is used to identify the next idle IU or the nearby RIS, until the information reaches $D$.

\begin{figure*}
    \centering
    \includegraphics[width=0.84\linewidth]{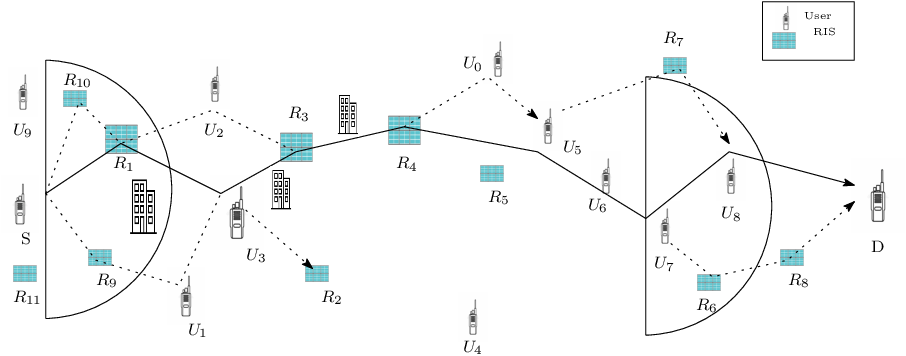}
    \vspace{-1mm}
    \caption{Illustrative Example}
    \label{pstrat}
    \vspace{-4mm}
\end{figure*}
\subsection{The Entire Procedure}

The complete process of information transfer, as described above, from $S$ to $D$ within a time limit of $T_d$ is the proposed framework DRAMS and this has been pictorially presented by a concise and compact flowchart in Fig. \ref{flowchart}.

The overall procedure is demonstrated by an illustrative example in Fig. \ref{pstrat}, where there is no direct LoS link between $S$ and $D$ and the corresponding delay constraint for the complete data transfer is $T_d$. To select the intermediate hops (IU/RIS), we consider a semi-circle of radius $r$ at $S$ in the LRD direction from $S$ to $D$. We observe from the figure that there are no available IUs within the half-circle $C_1$, which is centered at $S$ and satisfies the conditions as stated in \eqref{ykq}. Hence, $S$ estimates the waiting time interval of $\eta_w$ slots and the delay constraint $T_d$ is accordingly updated as $T_d^{'}=T_d- \eta_w T_s$. After waiting for $\eta_w$ slots, $S$ again scans the semi-circle of radius $r$ but still it cannot find an appropriate IU that can act as a relay. Accordingly, $S$ does not wait any further but observes that RIS $R_1,R_9$ and $R_{10}$ can act as potential reflectors for the first hop. As the reflected signal, via $R_1$, traverses the maximum distance in the LRD direction, $R_1$ is selected among the three. Since $r$ is the coverage distance for each RIS and idle $U_2,U_3$ lies inside this region, \eqref{optkq} is solved over these two IUs to select $U_3$ as the next hop.

By adopting a similar technique, we consider a right half circle of radius $r$ at $U_3$ to observe that there are no idle IUs in this region and $R_3$ is the only RIS that reduces the LRD from $U_3$ to $D$. Moreover, $R_4$ is selected to act as the next hop of this framework due to two reasons: firstly, there are no idle IUs available in the concerned region and secondly, as stated earlier, double reflections are non-negligible in practice. As $U_5$ covers the LRD towards $D$ and three consecutive RIS selection results in significant signal degradation \cite{2hop}, $U_5$ is chosen to act as the next relay node. In a similar logic, $U_7$ and $U_9$ are selected to act as the corresponding hops of the proposed framework. Therefore, the signal from $S$ reaches $D$ within a time limit of $T_d$ by using the path $S,R_1,U_3,R_3,R_4,U_5,U_7,U_9,D$.

\section{Delay Analysis}  \label{DA}

As stated earlier in Section III-C, here we investigate DRAMS in terms of the delay for information transfer from $S$ to $D$. Based on the choice of IUs and RISs as hops of this framework, we can have the following scenarios: (i) only IUs (both busy and idle) and (ii) both IUs and RISs are being used. Now we look at all the scenarios in detail.

\subsection{Only IUs}
Here we observe that the information packets from $S$ reach $D$ through a finite number of busy/idle IUs and no RIS is being used in this scenario. On arrival of information, $S$ immediately locates an idle IU $U_1$ within the right half circle of radius $r$, which is also in the LRD direction and can act as a DF relay. However, if $S$ cannot locate an idle IU in the desired region that satisfies the constraints from \eqref{ykq}, before proceeding with a RIS, $S$ waits for a finite amount of time to identify the next node in the proposed framework. We assume that there must be an IU that meets the constraints in \eqref{ykq} within this waiting time. In this context, we estimate the maximum acceptable waiting time $T_{d_i}$ at IU $U_i$, such that the overall delay constraint of time $T_d$ is not violated.

Let $l$ be the Euclidean distance from $S$ to $D$ and any IU can transmit to a maximum distance $r$. Hence, the minimum number of hops required to send a data packet from $S$ to $D$ is $\Psi=\displaystyle\biggl \lceil \frac{l}{r} \biggr \rceil$. Therefore, the maximum acceptable delay and actual delay at $S$ is $T_{d_0}=\displaystyle\frac{T_d}{\Psi}$ and $t_0$, respectively. In case the actual waiting time $t_i$ at $U_i$ is less than $T_{d_i}$, we propose that the leftover waiting time $T_{d_i}-t_i$ is carried forward to $U_{i+1}$, i.e., the maximum acceptable delay $T_{d_{i+1}}$ at $U_{i+1}$ is now updated as $T_{d_{i+1}}+(T_{d_i}-t_i)$.
\begin{rem}
    In this work, we consider a scenario, where IUs cannot communicate beyond a distance $r$, i.e., the IUs do not have a global knowledge of the system topology. Therefore, $U_i$ passes on $T_{d_i}-t_i$ to $U_{i+1}$, as it is unaware of exactly how many hops will be required in DRAMS for the complete information transfer from $S$ to $D$. 
\end{rem}

\noindent Accordingly, we characterize $T_{d_i}$ as
\begin{equation*}
    \displaystyle{T_{d_i}=\frac{T_d-\sum\limits_{k=0}^{i-2}\beta_kt_k-T_{d_{i-1}}}{\Psi-\Psi_i}+(T_{d_{i-1}}-\beta_{i-1}t_{i-1})} \quad i\geq 2,
\end{equation*}
\vspace{-1.2mm}
where $\Psi_i=\displaystyle \biggl \lfloor \frac{||S-U_i||}{r} \biggr \rfloor$, $\lfloor \cdot \rfloor$ is the floor function, $T_{d_0}=\dfrac{T_d}{\Psi}$, $T_{d_1}=\dfrac{T_d-T_{d_0}}{\Psi-\Psi_1}+(T_{d_0}-\beta_0t_0)$, and
\vspace{-1mm}
\begin{align}  \label{v2n}
\beta_i=\begin{cases} 
0, & \text{Idle} \:\: U_i ,\\
1, & \text{Busy}  \:\: U_i.
\end{cases}
\end{align}
Moreover, based on the value of $\beta$ corresponding to a particular $U_i$, we can have the following  extreme cases:
\begin{enumerate}
    \item $\beta_z=0$ $\forall$ $z<i$, i.e., no waiting at $U_1,\cdots,U_{i-1}$ and
    \item $\beta_z=1$ $\forall$ $z<i$, i.e., waiting at $U_1,\cdots,U_{i-1}$.
\end{enumerate}
Both these cases are investigated below.

\begin{table*} [!t] 
\begin{center}
  \caption{Summary of results.}
\resizebox{0.88\textwidth}{!}{%
{\renewcommand{\arraystretch}{2} 
  \begin{NiceTabular}{|c||c|c|c|}[hvlines,cell-space-limits=3pt] 
    \hline
    \bf{Cases} & \bf{IU} & \bf{RIS} & \textbf{Maximum acceptable delay at $U_i$}\\
    \hline\hline
    1 &  \checkmark & $\times$ & $\begin{aligned}
T_{d_i}=\begin{cases} 
\dfrac{T_d}{\Psi}, & i=0 ,\\[1ex]
\dfrac{T_d-T_{d_0}}{\Psi-\Psi_1}+(T_{d_0}-\beta_0t_0), & i=1,\\[1ex]
\dfrac{T_d-\sum\limits_{k=0}^{i-2}\beta_kt_k-T_{d_{i-1}}}{\Psi-\Psi_i}+(T_{d_{i-1}}-\beta_{i-1}t_{i-1}) & i\geq 2.
\end{cases}
    \end{aligned}$\\
    \hline
    2 & \checkmark & \checkmark & $\begin{aligned}
T_{d_i}=\begin{cases} 
\dfrac{T_d}{\Psi}, & i=0 ,\\[1ex]
\dfrac{T_d-T_{d_0}}{\Psi-\Psi_1} +(T_{d_0}-(c_0t'_0+(1-c_0) t''_0)), & i=1,\\[1ex]
\dfrac{T_d-\sum\limits_{k=0}^{i-2} c_kt'_k-\sum\limits_{k=0}^{i-2} (1-c_k)t''_k-T_{d_{i-1}}}{\Psi-\Psi_i} +(T_{d_{i-1}}-(c_{i-1}t'_{i-1}+(1-c_{i-1}) t''_{i-1})) & i\geq 2.
\end{cases}
    \end{aligned}$\\
\hline
  \end{NiceTabular}
  }
  }
  \label{tab:summ}
\end{center}
\vspace{-4mm}
\end{table*}

\subsection*{Case I: $\beta_z=0$ $\forall$ $z<i$.}
Here we investigate the scenario when the information from $S$ has not faced any waiting at $U_1,\cdots,U_{i-2},U_{i-1}$ till $U_i$. Accordingly, the maximum acceptable delay at $U_i$ is given by the following theorem.
\begin{theo}  \label{th1}
Without any delay in information transfer from $S$ through the IUs $U_1,\cdots,U_{i-2},U_{i-1}$, the maximum acceptable delay at $U_i$ is given by
\vspace{-1mm}
\begin{align} \label{b0f}
    T_{d_i}&= T_d \left( \sum\limits_{p=1}^i  \left(\frac{1}{\Psi-\Psi_p} \right) \prod\limits_{q=p+1}^i  \left( 1-\frac{1}{\Psi-\Psi_q} \right)
 \right. \nonumber \\
    & \qquad\qquad + \left.\frac{1}{\Psi}\prod\limits_{n=1}^i \left( 1-\frac{1}{\Psi-\Psi_n} \right) \right).
\end{align}
\end{theo}
\begin{proof}
    See Appendix \ref{app1}.
\end{proof}

We observe from Theorem \ref{th1} that the maximum acceptable delay $T_{d_i}$ at IU $U_i$ is expressed in terms of the overall delay constraint $T_d$. Furthermore, it can also be observed that $T_{d_i}$ increases monotonically with  $i$, i.e., $T_{d_i} \leq T_{d_{i+1}}$ $\forall$ $i$.

\subsection*{Case II: $\beta_z=1$ $\forall$ $z<i$.}
This implies that the information from $S$ has suffered delay at all the IUs till $U_i$, i.e., in this case, we obtain the maximum acceptable delay at $U_i$ from \eqref{v2n} as
\vspace{-1mm}
\begin{equation}  \label{b11}
    \displaystyle{T_{d_i}=\frac{T_d-\sum\limits_{k=0}^{i-2}t_k-T_{d_{i-1}}}{\Psi-\Psi_i}+(T_{d_{i-1}}-t_{i-1})} \quad i\geq 2,
\end{equation}
where $T_{d_0}=\dfrac{T_d}{\Psi}$ and $T_{d_1}=\dfrac{T_d-T_{d_0}}{\Psi-\Psi_1}+(T_{d_0}-t_0)$.

\noindent By simplifying $T_{d_1}$ in \eqref{b11}, we obtain
\vspace{-1mm}
\begin{align}  \label{td1b1}
    T_{d_1}& \overset{(a)}{=} T_d \left( \frac{1}{\Psi-\Psi_1} \left( 1-\frac{1}{\Psi} \right) + \frac{1}{\Psi} \right) \nonumber \\
    & \qquad -\mu_1 \ln \left( \left(\frac{1-e^{-T_s/\mu_1}}{p_{\rm th}} \right)^+\right) -T_s \nonumber \\& \!\!\!\!\!\!\!\!\!\!\!\!\!\!\!\!\!\!\!\!\!\!\!\! \overset{(b)}{=} T_d \left( \frac{1}{\Psi-\Psi_1} \left( 1-\frac{1}{\Psi} \right) + \frac{1}{\Psi} \right) -\mu_1 \ln \left( \left(\frac{T_s}{\mu_1p_{\rm th}} \right)\right) -T_s,
\end{align}
where (a) follows from $T_{d_0}=\dfrac{T_d}{\Psi}$ and \eqref{iwait}. Furthermore, (b) follows from the Taylor expansion of $e^{-T_s/\mu_1}$ and neglecting its higher order terms as we know from Section \ref{tcharc} that $T_s/\mu_1<1$ and finally, by considering an appropriate $p_{\rm th}$ such that $\frac{T_s}{\mu_1p_{\rm th}}>1$ holds.

\begin{rem}  \label{rmb1}
    Note that for $x \geq 1$, $x\ln \left( \dfrac{1}{x} \right)$ is a monotonically decreasing function with its maxima at $x=1$, when $x\ln \left( \dfrac{1}{x} \right)=0$. Hence, we can observe from \eqref{td1b1}, that $T_{d_1}$ increases with $\mu_1$ when other parameters are constant. It is interesting to observe, that the same intuition is also provided by the term $\text{p}_{10}$ in the transition probability matrix stated in \eqref{tpmtrx}.
\end{rem}

Instead of making a claim specifically with respect to $U_1$ as in Remark \ref{rmb1}, we can make a generalization as follows. From \eqref{b11}, we obtain
\vspace{-1mm}
\begin{align}  \label{tdi}
    T_{d_i}&=\dfrac{T_d-T_{d_{i-1}}}{\Psi-\Psi_i}+T_{d_{i-1}}-\dfrac{\sum\limits_{k=0}^{i-2}t_k}{\Psi-\Psi_i}-t_{i-1}\qquad i\geq 2 \nonumber \\
    &=\dfrac{T_d-T_{d_{i-1}}}{\Psi-\Psi_i}+T_{d_{i-1}}- \left( \frac{i-1}{\Psi-\Psi_i}+1 \right)T_s \nonumber \\
    & \!\!\!\!\!\! - \underbrace{\left( \frac{1}{\Psi-\Psi_i} \sum\limits_{k=1}^{i-1} \mu_k \ln \left( \frac{T_s}{\mu_kp_{\rm th}}  \right) +\mu_i \ln \left( \frac{T_s}{\mu_ip_{\rm th}}  \right) \right)}_{\text{function of} \:\:\mu_1,\cdots,\mu_i},
\end{align}
which is based on \eqref{iwait}.

\begin{rem}
We observe from \eqref{tdi} that $T_{d_i}$ is a joint function of $\mu_1,\cdots,\mu_i$. This implies that the average acceptable time delay at an arbitrary intermediate user is dependent on the traffic characteristics of all the previous intermediate users.
\end{rem}

Since we have investigated the two extreme scenarios, i.e., no delay at $U_1,\cdots,U_{i-1}$ and waiting at all of $U_1,\cdots,U_{i-1}$, the actual $T_{d_i}$ corresponding to $U_i$ will be in between the two. The reason behind this observation is attributed to the practical scenario of $\beta=0$ for some of the IUs and $\beta=1$ otherwise.

\subsection{RIS and IUs}
This is the most general scenario, which involves IUs, both idle and busy, and RISs in the process of information transfer from $S$ to $D$. In this setup, if $S$ has information packets to transfer, first it will search for idle IUs within the right half circle of radius $r$ in the LRD direction. If one of the IUs is available that meets all the criteria as stated in \eqref{ykq} within the acceptable waiting time, this particular IU will serve as the DF relay node but otherwise, $S$ goes for a RIS. This process of IU or RIS selection is repeated at each hop until $D$ is reached. In this context, we come across namely two types of delays at an arbitrary IU $U_i$ as follows.
\begin{enumerate}[(i)]
    \item Delay $t_i'$: waiting time at $U_i$, when it transfers the data packets to the following IU and
    \item Delay $t_i''$: time after which $U_i$ chooses an RIS to do the transfer due to reasons such as the unavailability of an appropriate IU. 
\end{enumerate}
It is to be noted that both $t_i'$ and $t_i''$ occur at $U_i$ for information transfer to $U_{i+1}$. Towards this direction, $U_i$ after suffering a delay $t_i''$, immediately chooses a RIS and if there are no suitable IU in the LRD direction of the RIS, it immediately directs the signal to its adjacent RIS. This is based on the fact that double-RIS-aided information transfer is a non-negligible phenomenon \cite{2hop} and the corresponding channel model is described in Section \ref{cmod}. Moreover, we assume that the RISs are strategically placed \cite{deb2021ris} and as a result, there is always another RIS in the right half circle of radius $r$. Furthermore, in the context of maximum acceptable delay calculation at an IU, here also we propose $T_{d_i}-t_i$ of being carried forward to only $U_{i+1}$.

With the motivation and the framework for this variant already stated in the previous section, we proceed along similar lines to obtain the maximum acceptable delay $T_{d_i}$ corresponding to $U_i$.
\vspace{-1mm}
\begin{align}
    T_{d_i}&=\dfrac{T_d-\sum\limits_{k=0}^{i-2} c_kt'_k-\sum\limits_{k=0}^{i-2} (1-c_k)t''_k-T_{d_{i-1}}}{\Psi-\Psi_i} \nonumber \\
    & \!\!\!\! +(T_{d_{i-1}}-(c_{i-1}t'_{i-1}+(1-c_{i-1}) t''_{i-1})) \qquad i \geq 2,
\end{align}
and $T_{d_1}=\dfrac{T_d-T_{d_0}}{\Psi-\Psi_1} +(T_{d_0}-(c_0t'_0+(1-c_0) t''_0))$, where
\vspace{-1mm}
\begin{align}  \label{prop1}
c_i=\begin{cases} 
1, & \text{for D2D delay} ,\\
0, & \text{else}
\end{cases}
\end{align}
Note that for a particular $U_i$, both $t_i'$ and $t_i''$ cannot exist at the same time.

Hence in this section, we have analyzed the transmission delay for all the possible cases of the proposed framework. Finally, Table \ref{tab:summ} presents a summary of the main analytical results derived in this section.

\section{Numerical Results}
In this section, we carry out extensive simulations to validate the performance of DRAMS and also compare with the nearest existing approach. Here we consider data transmission in a Rician fading scenario, where we assume Rician factor $K=10$ dB. The default parameters considered are: slot duration $T_s=100$ $\mu$s \cite{crn}, IU transmission power $P=30$ dBm, IU processing power $P_{\rm proc}=10$ dBm, pathloss at one-meter distance $\rho_{\rm L}=10^{-3.53}$ \cite{prmtr1}, pathloss exponent $\alpha=4.2$ between two consecutive IUs and $\alpha=2$ elsewhere, the number of elements in each RIS $N=250$, and acceptable delay bound $T_d=50$ ms \cite{prmtr2}. The parameters related to energy harvesting at the IUs are $M=24$ mW, $a=150,$ and $b=0.014$ \cite{npara}. Based on \eqref{mqam}, we consider M-ary quadrature amplitude modulation (M-QAM) transmission between consecutive IUs and a BER of $P_b=10^{-6}$ results in $\dfrac{P\rho_{\rm L}d^{-\alpha}|h|^2}{\sigma^2}=9.6724(m_q-1)$, where $m_q$ is constellation size as defined in Section \ref{modulation}. Accordingly, we obtain the transmission modes (TM) as stated in Table \ref{tabrst} and these TM are used in this section.
\begin{table} [!t]
\centering
  \caption{Transmission Modes for $P_b=10^{-6}$.} \label{tabrst}
\resizebox{\columnwidth}{!}{%
  \begin{tabular}{|c|c|c|}
    \hline \hline
    \textbf{SNR interval (dB)} & \textbf{Modulation} &  \textbf{Rate (bits/sym.)}\\
    \hline
     $(-\infty, 9.8554)$ & No transmission & 0 \\
    \hline
   $[9.8554, 12.8657)$ & BPSK & 1 \\
    \hline
    $[12.8657,14.6266)$ & QPSK & 2 \\
    \hline
    $[14.6266,15.8760)$ & $8$-QAM & 3 \\
    \hline
	$[15.8760,16.8451)$ & $16$-QAM & 4 \\
    \hline
    $[16.8451,17.6369)$ & $32$-QAM & 5 \\
    \hline
	$[17.6369,18.3063)$ & $64$-QAM & 6 \\   
    \hline
    $[18.3063, 18.8863)$ & $128$-QAM & 7 \\
    \hline
    $[18.8863, +\infty)$ & $256$-QAM & 8 \\
    \hline
  \end{tabular}
  }
  \vspace{-3.2mm}
\end{table}
Moreover, in this work, when an RIS is chosen due to the unavailability of IUs, we consider parameters $M_b=1000$ and $\varepsilon=10^{-4}$ \cite{shanfp}. Next we demonstrate the performance of DRAMS and also validate the proposed analytical framework against Monte Carlo simulation. Finally, we compare DRAMS with the existing approaches.

\subsection{Performance of DRAMS}

\begin{figure}[!t]
\centering\includegraphics[width=\linewidth]{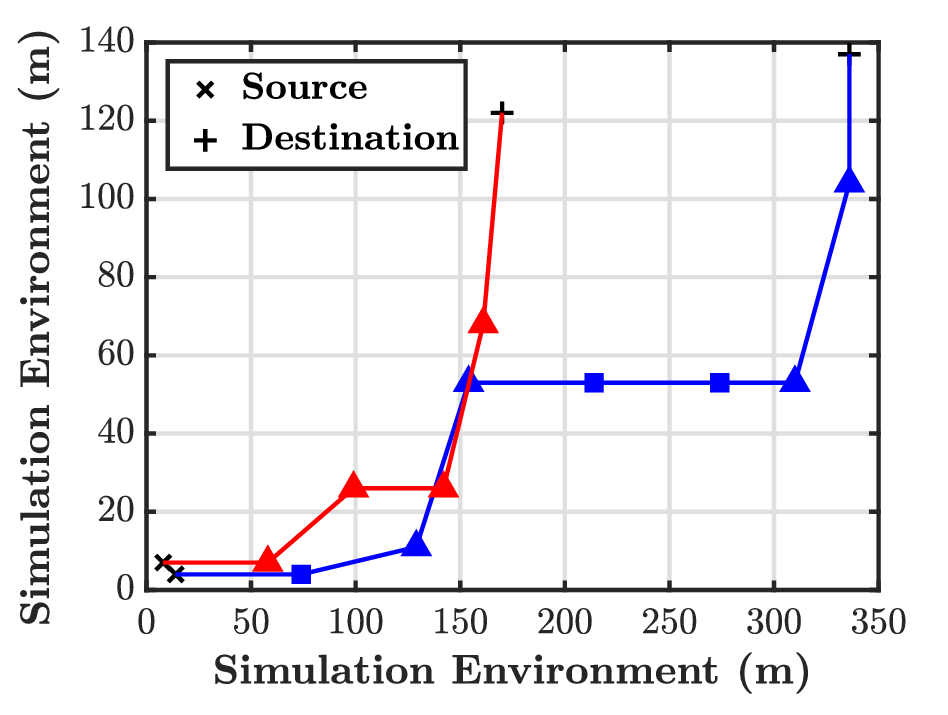}
\caption{DRAMS trajectories for two different scenarios; $\Delta$ corresponds to IU and $\Box$ corresponds to RIS.}
\label{fig:traj}
\vspace{-4mm}
\end{figure}

An illustration of the DRAMS-based multihop trajectory is presented in Fig. \ref{fig:traj}. In this scenario, we consider a two-dimensional squared area of $400 \times 400$ $\rm m^2$, where the IUs and the RISs are randomly and strategically placed, respectively, as shown in Fig. \ref{fig:traj}. Moreover, we assume an IU coverage of $60$ m, where `coverage' refers to the maximum distance at which a particular IU can communicate. We consider two separate instances, when a randomly selected $S-D$ pair wants to communicate. In the process of doing so, DRAMS establishes a multi-hop connection, which effectively brings out the advantages of the proposed scheme as follows.
\begin{enumerate}
    \item IUs are preferred over RISs in establishing the $S-D$ connection, i.e., the figure demonstrates that if IUs are available to act as relays without violating the delay constraint, RISs are completely overlooked by DRAMS.
    \item On the other hand, the RISs are considered as an option only in the case of IU unavailability. Furthermore, to take advantage of the RISs significantly, DRAMS also leverages on the secondary reflections from the RISs, which can also be observed by the choice of two consecutive $\Box$s in the figure.
\end{enumerate}
In this way, by prioritizing the choice of IUs over RISs, DRAMS avoids the aspect of unnecessary resource wastage. However, it is to be noted that this benefit does not come at a cost of violating the delay constraint and the corresponding analysis is already explained in Section \ref{DA}.

\begin{figure}[!t]
\centering\includegraphics[width=\linewidth]{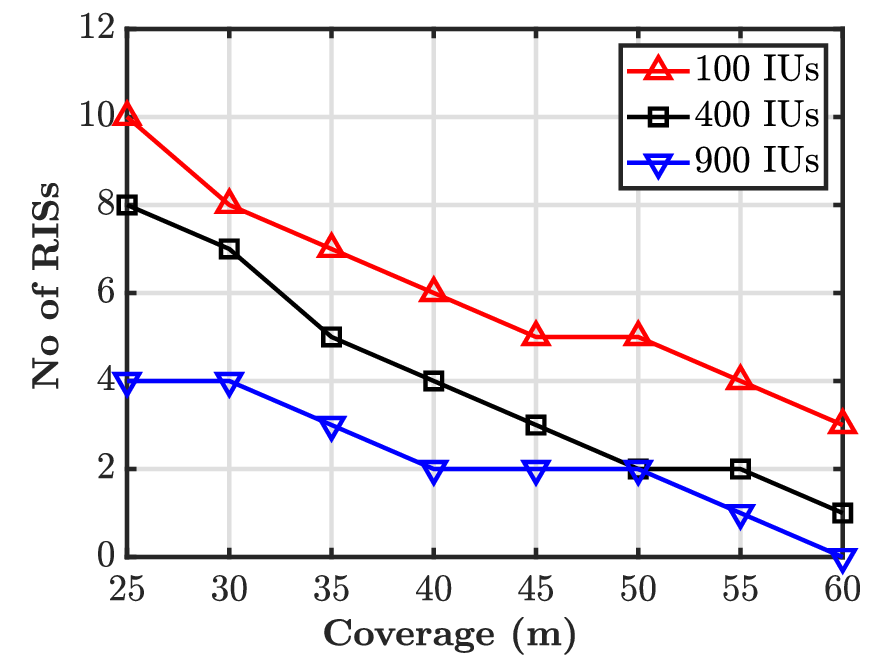}
\caption{Effect of IU coverage.}
\label{fig:cov}
\vspace{-4mm}
\end{figure}

Fig. \ref{fig:cov} illustrates the number of RISs used to connect $S$ to $D$ as a function of the IU coverage. We consider a particular $S-D$ pair and three different scenarios with IU density $100,400,$ and $900$, respectively. Accordingly, we look at the number of RISs used to establish a multi-hop connection from $S$ to $D$. For a particular IU density, we observe that the number of RISs used to connect $S-D$ follows a non-increasing trend with respect to increasing IU coverage. This is justified by the fact that a smaller coverage implies more number of hops to connect $S-D$. Moreover, we know that higher carrier frequency results in higher data rate but lower coverage. Hence, the figure demonstrates that for identical IU density, higher carrier frequency (lower coverage) results in higher number of RISs being used and vice-versa. Finally, we also note that irrespective of the IU density, the number of RISs used asymptotically reaches zero with increasing  IU coverage.

\begin{figure}[!t]
\centering\includegraphics[width=\linewidth]{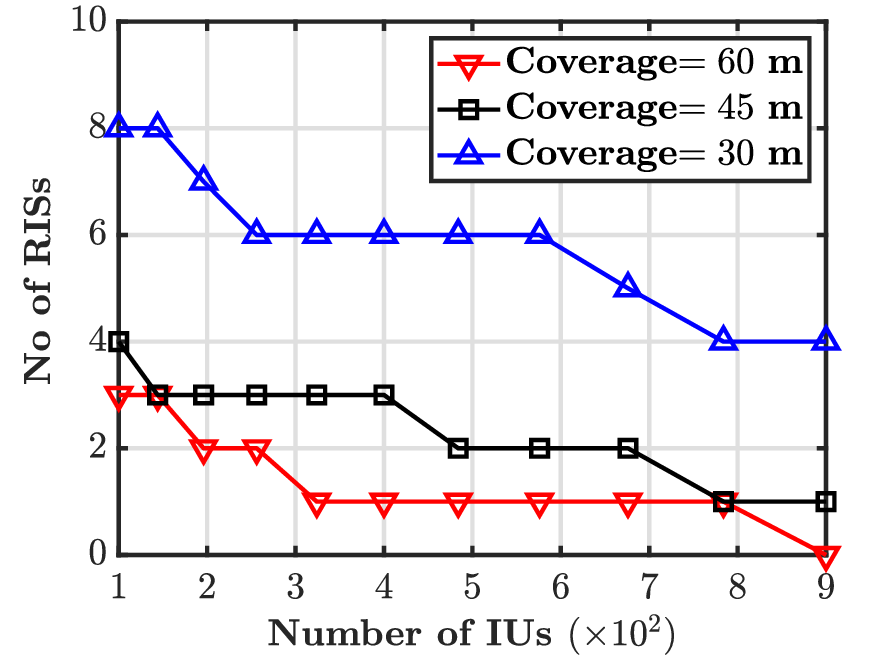}
\caption{Effect of IU density.}
\label{fig:iu}
\vspace{-4mm}
\end{figure}

Fig. \ref{fig:iu} depicts the number of RISs used to connect $S$ to $D$ as a function of the IU density. In this figure, we establish a connection between a $S-D$ pair for three different IU coverage of $30,45,$ and $60$ m, respectively. It is observed here that a lesser number of RISs are being used as the IU density increases and here also, irrespective of the IU coverage radius, the value asymptotically reaches zero. In other words, for an environment with a significantly large IU density, it is possible to completely avoid the usage of RISs.

From Fig. \ref{fig:cov} and Fig. \ref{fig:iu} we observe that depending on the carrier frequency (i.e., the IU coverage) and density of IUs in the surroundings, it is possible to establish a multihop $S-D$ connection consisting of only IUs and not RISs. This further strengthens our claim of exploiting the IU traffic characteristics to reduce the dependency on the RISs. It is to be noted that our proposed DRAMS avoids wastage of resources but not at the cost of performance degradation.

\subsection{Verification of $\nu_B$ and $\nu_I$  by Monte Carlo Simulation}

\begin{figure}[!t]
\centering\includegraphics[width=\linewidth]{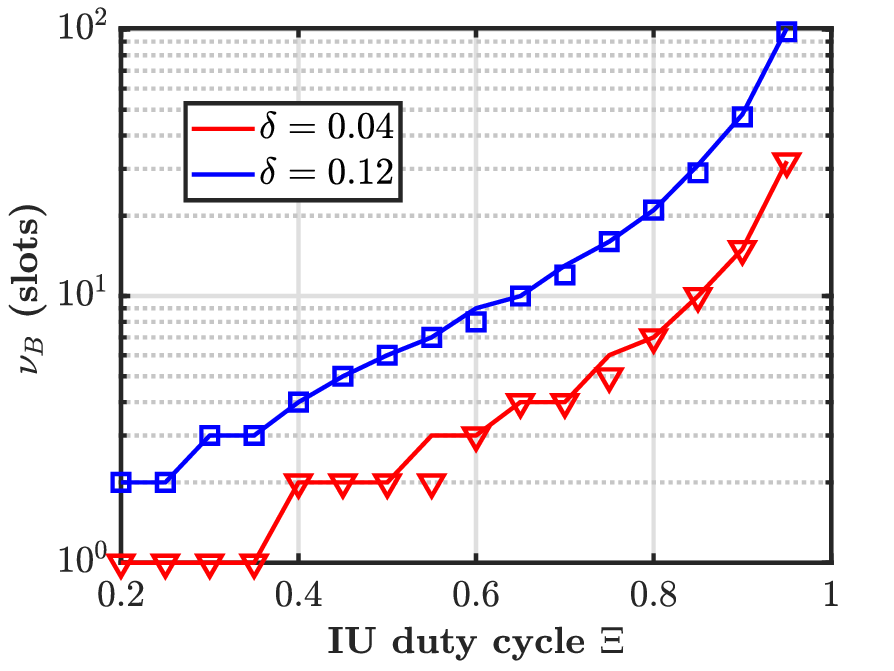}
\caption{Verification of $\nu_B$ estimation; lines correspond to analysis and markers correspond to simulation results.}
\vspace{-4mm}
\label{fig:zb}
\end{figure}

Here for the generation of results, we define the average IU activity duty cycle as $\Xi=\dfrac{\mu_k}{\mu_k+\lambda_k}$ $\forall \:\: k$, which is the average fraction of the total time that an IU remains busy communicating with its own dedicated receiver.

Fig. \ref{fig:zb} compares the analytically obtained $\nu_B$ in (\ref{zb}) with the Monte Carlo simulations, where we consider the average `OFF' duration $\lambda_k=4$ ms $\forall$ $k$. It is observed that $\nu_B$ increases monotonically with $\Xi$ and moreover, the rate of increase exponentially shoots up as $\Xi \rightarrow 1$. This is also intuitive, as increasing $\Xi$ implies that the IU will remain busy most of the time and hence, the time duration for which it will remain busy given that it is currently busy will also increase. Furthermore, we also observe that for a particular $\Xi$, a higher value of $\delta$ implies a greater value of $\nu_B$ and vice-versa, as evident from \eqref{zb}.

\begin{figure}[!t]
\centering\includegraphics[width=\linewidth]{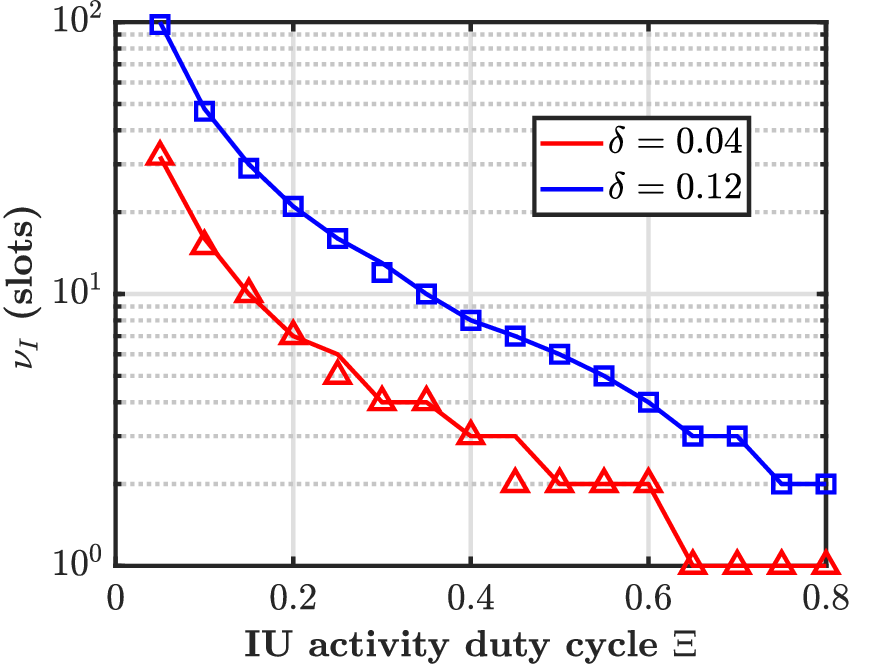}
\caption{Verification of $\nu_I$ estimation; lines correspond to analysis and markers correspond to simulation results.}
\vspace{-4mm}
\label{fig:zi}
\end{figure}

We compare the analytically obtained $\nu_I$ with the corresponding Monte Carlo simulations in Fig. \ref{fig:zi}, where we consider the average `ON' duration $\mu_k=4$ ms $\forall$ $k$. We observe that, irrespective of $\delta$, the value of $\nu_I$ decreases with increasing $\Xi$, unlike $\nu_B$. This is intuitive too, as increasing the duty cycle implies that the IU will remain idle for a relatively lesser amount of time. Furthermore, here also, we observe that for any particular $\Xi$, $\delta_1>\delta_2$ results in $\nu_I$ corresponding to $\delta_1$ being greater than the $\nu_I$ corresponding to $\delta_2$. Thus, based on Fig. \ref{fig:zb} and Fig. \ref{fig:zi}, we can state that $\nu_B$ and $\nu_I$ complement each other. Furthermore, it can also be said, that DRAMS will always have a tendency to select IUs with lower $\Xi$ as relays.

\subsection{Performance Comparison}

Here we first define the metrics, namely data throughput and energy efficiency, which will be used to quantify the performance of DRAMS.

\subsubsection{Data throughput $\mathcal{D_T}$}
By assuming that DRAMS chooses $\mathcal{X}$ IUs and a certain number of RISs to connect the $k$-th $S-D$ pair in $\mathcal{X}+1$ hops, the data throughput is defined as
\vspace{-1mm}
\begin{equation}
    \mathcal{D_T}=\dfrac{1}{\sum\limits_{i=1}^{\mathcal{X}+1} \dfrac{1-a_i}{\left( 1-P_b \right) m_{q_i}}+\dfrac{a_i}{R_i\left(\gamma_i\right)}},
\end{equation}
where
\vspace{-1mm}
\begin{align*}  \label{dr1}
a_i=\begin{cases} 
1, & \text{if $i$-th hop involves RIS} ,\\
0, & \text{else}.
\end{cases}
\end{align*}
Here $P_b$ is the BER and $m_{q_i}$ is the constellation size as stated in \eqref{mqam}. Moreover, in case we have a RIS selected due to idle IU unavailability, $R_i$ is the corresponding achievable data rate as defined in \eqref{rate}.

\begin{figure*}
\centering\includegraphics[width=0.8\linewidth]{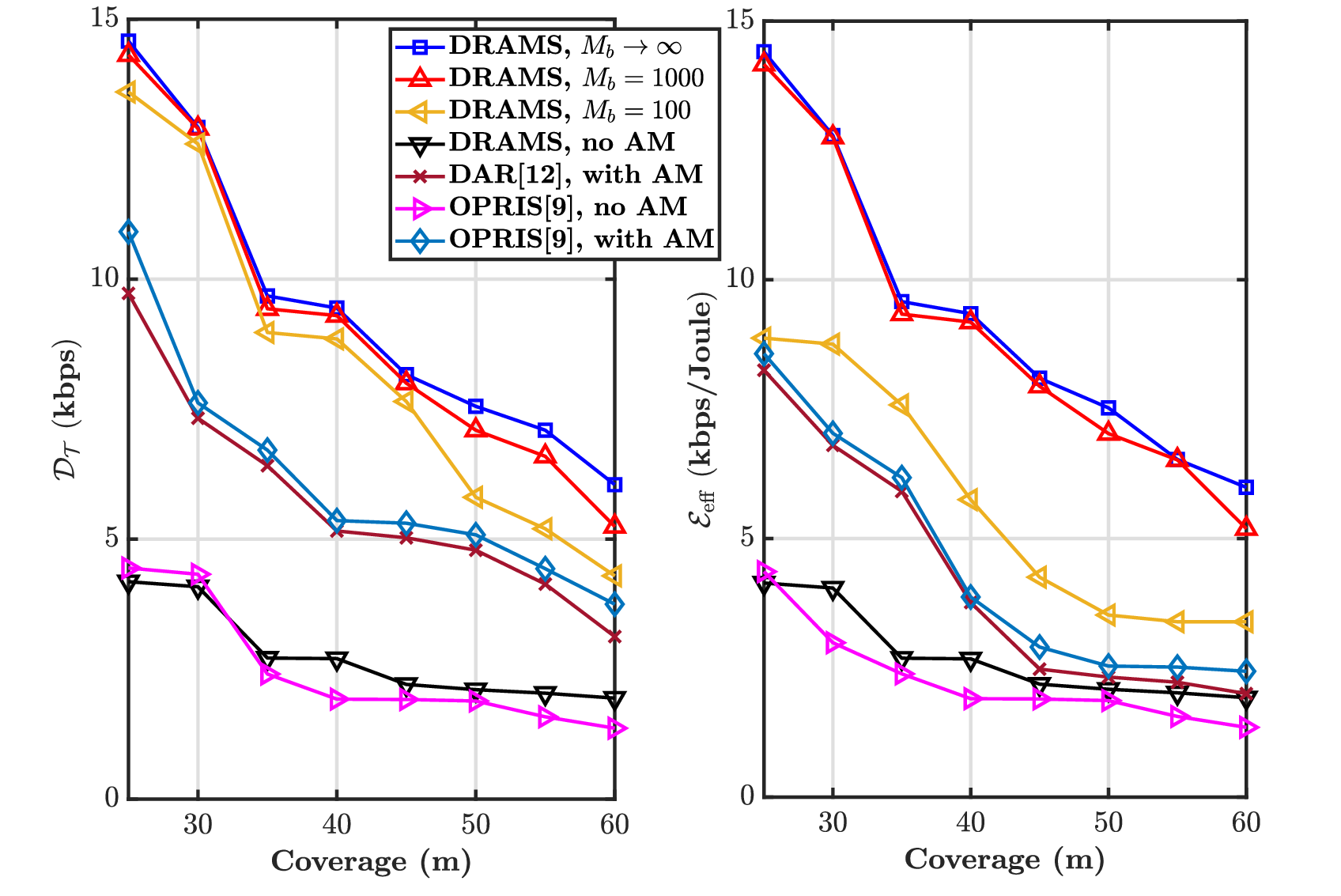}
\caption{Performance comparison: (a) Data throughput (b) Energy efficiency.}
\vspace{-4mm}
\label{fig:compare}
\end{figure*}

\subsubsection{Energy efficiency $\mathcal{E}_{\rm eff}$}
Similar to data throughput, the system energy efficiency for transferring $\alpha_k$ packets of data with $\varphi$ bits in each packet, when the $k$-th $S-D$ pair gets connected in $\mathcal{X}+1$ hops is defined as
\vspace{-1mm}
\begin{equation}  \label{ee}
    \mathcal{E}_{\rm eff}=\dfrac{\alpha_k\varphi}{\left( P+P_{\rm proc} \right)T_s\sum\limits_{i=1}^{\mathcal{X}+1}\tau_i-T_s\sum\limits_{j=1}^{\mathcal{X}}P_{\rm harv}(j)\tau_j},
\end{equation}
where $P$ and $P_{\rm proc}$ is the transmission and processing power respectively, as defined in \eqref{enrgy}. Moreover, $\tau_i$ is the time required (in slots) for complete information transfer in the $i$-th hop and finally, based on the channel condition, $P_{\rm harv}(j)$ is the harvested power at the $j$-th IU when the channel between the $(j-1)$-th and $j$-th IU is being used for information transfer. Note that in \eqref{ee}, the numerator is essentially the amount of data transferred and the denominator denotes the net energy consumption, where the first term is the energy required for data transmission and the second term implies the amount of energy harvested. Hereafter, we compare the performance of DRAMS with the existing benchmark schemes \cite{opris,dar}. Accordingly, the variants used for this purpose are the following:
\begin{enumerate}
    \item OPRIS \cite{opris}: The work investigates the optimal placement of RISs when the $S-D$ pair is either directly connected or connected via a single RIS only. In other words, it does not involve any aspect of connecting two devices that require multihop communication.
    \item DAR \cite{dar}: This work proposes the involvement of only double-RIS aided scenarios for the purpose of connecting $S$ to $D$. It does not involve the aspect of incorporating single and/or double reflection depending on the availability of IU and/or RIS.
\end{enumerate}

Fig. \ref{fig:compare} (a) demonstrates an overall decreasing trend of $\mathcal{D_T}$ with IU coverage, irrespective of the deployed scheme. This is because although increasing coverage implies lesser number of hops, the pathloss factor becomes dominant. It is observed that the performance of both DRAMS and OPRIS, without AM, are comparable and equally poor as compared to DRAMS irrespective of $M_b$. This degraded performance is attributed to the inability to exploit the temporal variation of the wireless channel. Moreover, we note that irrespective of coverage, DRAMS with $M_b \rightarrow \infty$ always outperforms its finite $M_b$ counterpart and the performance gap increases with decreasing $M_b$. This is because the $M_b \rightarrow \infty$ scenario always chooses the best channel in each hop while connecting $S-D$, whereas a finite $M_b$ scenario cannot do so always due to the application-specific delay constraints. As a result, it can be concluded that if delay is not a critical factor for the application at hand, the performance of DRAMS gets enhanced by a finite margin. It is important to note that OPRIS (DAR) always utilizes single (double) reflection in the RIS-aided networks. On the contrary, DRAMS uses either of the two, depending on the scenario, which results in reduced RIS deployment, but not at the cost of degraded data throughput. As in Fig. \ref{fig:compare} (a), Fig. \ref{fig:compare} (b) depicts the advantage of the proposed framework in terms of $\mathcal{E}_{\rm eff}$, which also reduces with increasing coverage irrespective of the framework being used.



\subsection{Impact of Mobility}

\begin{figure}[!t]
\centering\includegraphics[width=\linewidth]{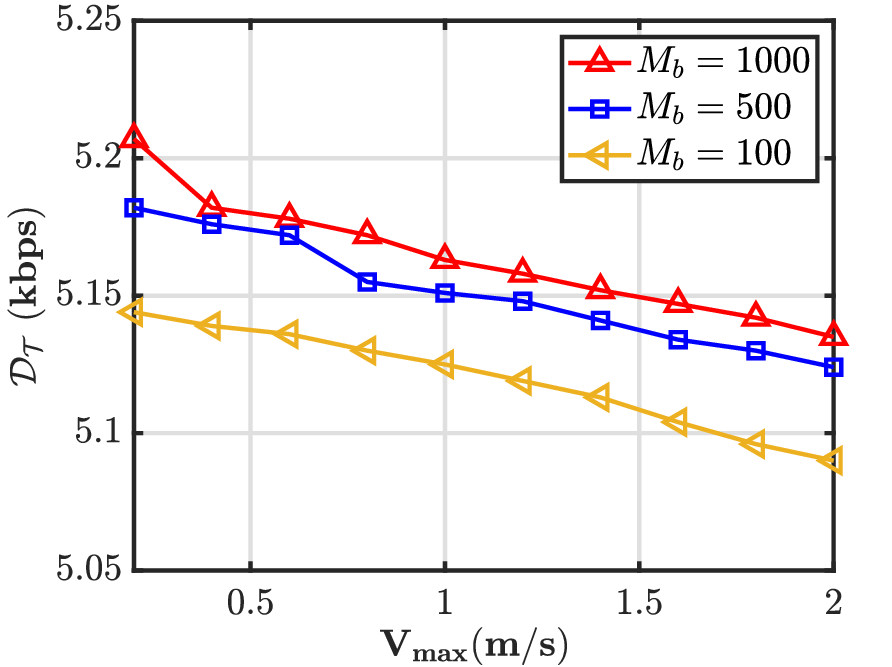}
\caption{Effect of IU mobility on DRAMS.}
\vspace{-4mm}
\label{fig:mob}
\end{figure}

Here we consider a mobile scenario, i.e., the IUs are having a particular velocity in a certain direction. Note that this will have a significant effect on the performance of the proposed scheme, as the inter-IU distance changes during the process of information transfer. In this context, Fig. \ref{fig:mob} illustrates the impact of IU mobility on DRAMS, with the coverage being taken as $50$ m and the mobility aspect being modeled by the standard random waypoint (RWP) model \cite{mobility}. Here, we fix the maximum possible velocity $V_{\max}$ of the IUs, consider a random velocity in $[0,V_{\max}]$ and observe its impact on the system data throughput. It can be seen that the performance deteriorates monotonically with increasing $V_{\max}$, which is intuitive.

The reason for this is attributed to the fact that, due to mobility, a particular IU can move outside the coverage of another IU even during the communication process. This inevitably leads to outage, resulting in lesser data rate. Moreover, note that this figure with $V_{\max}=0$ is a special case corresponding to the static scenario, i.e., Fig. \ref{fig:compare}. Furthermore, note that for a fixed $V_{\max}$, $\mathcal{D_T}$ decreases with $M_b$, which is inline with the observation made in Fig. \ref{fig:compare} as well.
 
\section{Conclusion}
In this paper we proposed a novel double-RIS assisted adaptive modulation-based multihop routing scheme for D2D wireless networks, which takes into account the aspect of multi-RIS secondary reflection. The proposed DRAMS exploits the traffic characteristics of the users present in the surroundings to bring down the dependency on the already deployed RISs, which reduces the wastage of resources. Numerical results demonstrate that double RIS assistance, at times, provide indirect LoS between the $S-D$ pair, which may not be achieved even with single RIS assistance. Moreover, we observe that, in certain cases, neighbouring IUs alone are sufficient for providing the required LoS without any RIS. Moreover, the results also showcase the significance of the proposed framework in terms of enhanced data throughput and energy efficiency. An immediate extension of this work is to investigate a non-cooperative scenario, where the users are independent to decide whether they would like to act as a relay and if they do, then for which corresponding $S-D$ pair in case of multiple requests.

\appendix

\section{Proof of Theorem \ref{th1}}  \label{app1}
By replacing $\beta_z=0$ $\forall$ $z<i$ in \eqref{v2n}, we obtain
\begin{equation} \label{v2n1}
    T_{d_i}=\dfrac{T_d-T_{d_{i-1}}}{\Psi-\Psi_i}+T_{d_{i-1}} \quad i\geq 1,
\end{equation}
where $T_{d_0}=\dfrac{T_d}{\Psi}$. After trivial manipulations, $T_{d_1}$ can be alternatively written as
\vspace{-1mm}
\begin{equation} \label{td1}
    T_{d_1}\!=\dfrac{T_d-T_{d_0}}{\Psi-\Psi_1}+T_{d_0} =\! T_d \left(\!\frac{1}{\Psi-\Psi_1}+ \frac{1}{\Psi}\left( 1-\frac{1}{\Psi-\Psi_1} \right)\!\right).
\end{equation}
Similarly, we obtain $T_{d_2}$ as a function of $T_{d_1}$, which in turn, can be further simplified in terms of $T_{d_0}$ as
\vspace{-1mm}
\begin{align}  \label{td2}
    T_{d_2}&=\frac{T_d}{\Psi-\Psi_2}+ \left( 1-\frac{1}{\Psi-\Psi_2} \right) T_{d_1} \nonumber \\
    &\overset{(a)}{=}\frac{T_d}{\Psi-\Psi_2}+\left( 1-\frac{1}{\Psi-\Psi_2} \right) \nonumber \\
    & \qquad \qquad \times \left(\frac{T_d}{\Psi-\Psi_1}+ \left( 1-\frac{1}{\Psi-\Psi_1} \right) T_{d_0} \right) \nonumber \\
    &=T_d\left( \frac{1}{\Psi-\Psi_2} + \frac{1}{\Psi-\Psi_1}\left( 1-\frac{1}{\Psi-\Psi_2} \right) \right. \nonumber \\
    & + \left. \dfrac{1}{\Psi} \left( 1-\frac{1}{\Psi-\Psi_1} \right) \left( 1-\frac{1}{\Psi-\Psi_2} \right) \right),
\end{align}
where (a) follows from \eqref{td1}. By proceeding in the same way for $i \geq 1$, we get
\vspace{-1mm}
\begin{align}  \label{th1m}
    T_{d_i}&= T_d \left( \sum\limits_{p=1}^i  \left(\frac{1}{\Psi-\Psi_p} \right) \prod\limits_{q=p+1}^i  \left( 1-\frac{1}{\Psi-\Psi_q} \right)
 \right. \nonumber \\
    & \qquad\qquad + \left.\frac{1}{\Psi}\prod\limits_{n=1}^i \left( 1-\frac{1}{\Psi-\Psi_n} \right) \right).
\end{align}
Moreover, it can also be observed that by putting $i=2$ in \eqref{th1m} results in \eqref{td2}.

\bibliographystyle{IEEEtran}
\bibliography{ref}

\end{document}